\documentstyle[prd,aps,epsf]{revtex}

\def\rn{}
\def\nn#1 #2{#2. #1}				
\def\nnn#1 #2 #3{#2. #3. #1}			
\def\nnnn#1 #2 #3 #4{#2. #3. #4 #1}		
\def\nnnnn#1 #2 #3 #4 #5{#2. #3. #4 #5. #1}	
\def\dualand{ and\hbox{ }}				
\def\multiand{, and\hbox{ }}				
\def\rf#1;#2;#3;#4;#5 {{\frenchspacing\par\rn#1, #3 {\bf #4}, #5 (#2). \par}}
\def\rg#1;#2;#3;#4;#5;#6 {{\frenchspacing\par\rn#1, #3 {\bf #4}, #5 (#2). \par}}
\def\rfbook#1;#2;#3;#4;#5 {{\frenchspacing\par\rn#1, {\it #3} (#5, #4, #2).\par}}
\def\rfprep#1;#2;#3 {{\par\frenchspacing\rn#1, #3 (#2).\par}}
\def\rfproc#1;#2;#3;#4;#5;#6 {{\frenchspacing\par\rn#1 #2, in {\it #3}, ed. #4 (#5: #6)\par}}
\def\rfprocp#1;#2;#3;#4;#5;#6;#7 {{\frenchspacing\par\rn#1 #2, in {\it #3}, ed. #4 (#5: #6), p#7\par}}

\def\beql#1{\begin{equation}\label{#1}}
\def\beqa#1{\begin{eqnarray}\label{#1}}
\def\eeqa{\end{eqnarray}}

\def\spose#1{\hbox to 0pt{#1\hss}}
\def\simlt{\mathrel{\spose{\lower 3pt\hbox{$\mathchar"218$}}
     \raise 2.0pt\hbox{$\mathchar"13C$}}}
\def\simgt{\mathrel{\spose{\lower 3pt\hbox{$\mathchar"218$}}
     \raise 2.0pt\hbox{$\mathchar"13E$}}}
\def\simpropto{\mathrel{\spose{\lower 3pt\hbox{$\mathchar"218$}}
     \raise 2.0pt\hbox{$\propto$}}}

\def\etal{{\frenchspacing\it et al.}}

\def\fn{f_\nu}

\def\erfc{{\rm erfc}\,}

\newcommand{\beq}{\begin{equation}}
\newcommand{\eeq}{\end{equation}}

\def\lap{\lower.5ex\hbox{$\; \buildrel < \over \sim \;$}}
\def\gap{\lower.5ex\hbox{$\; \buildrel > \over \sim \;$}}

\def\L{\Lambda}

\def\rL{\rho_\L}

\def\be{\begin{equation}}
\def\ee{\end{equation}}
\def\ba{\begin{eqnarray}}
\def\ea{\end{eqnarray}}

\begin{document}

\title{Anthropic predictions for vacuum energy and neutrino masses} 

\author{Levon Pogosian$^1$, Alexander Vilenkin$^1$, and Max Tegmark$^2$}

\address{
$^1$ Institute of Cosmology, Department of Physics and Astronomy,\\
Tufts University, Medford, MA 02155, USA \\
$^2$ Department of Physics and Astronomy, University of Pennsylvania,\\
Philadelphia, PA 19104, USA}

\wideabs{
\maketitle

\begin{abstract}

It is argued that the observed vacuum energy density and the small
values of the neutrino masses could be due to anthropic selection
effects. Until now,
these two quantities have been treated separately
from each other and, in particular,  anthropic predictions for the vacuum energy 
were made under the assumption of zero neutrino masses. Here we consider
two cases. In the first, we
calculate predictions for the vacuum energy for a fixed (generally
non-zero) value of the neutrino mass. In the second we
allow both quantities to
vary from one part of the universe to another. We find that the
anthropic predictions for the vacuum energy density are in a better
agreement with observations when one allows for non-zero neutrino
masses. We also find that the individual distributions
for the vacuum energy and the neutrino masses are reasonably robust and do
not change drastically when one adds the other variable.

\end{abstract}
\pacs{} }
\narrowtext

\section{Introduction}

The smallness of the vacuum energy density $\rL$ (the cosmological
constant) is one of the most perplexing mysteries that we now face in
theoretical physics.  The values of $\rL$ indicated by particle
physics models are far too large, at least by 60 orders of magnitude.
The only plausible explanation that has so far been suggested is that
$\rL$ is a stochastic variable, taking different values in different
parts of the Universe, and that the smallness of the observed value is
due to anthropic selection effects. The proposed selection mechanism
is very simple\footnote{For earlier suggestions that the cosmological
constant may be anthropically selected, see
\cite{Davies,BT,LindeLambda}.}
\cite{Weinberg87,Vilenkin95a,Efstathiou95,Vilenkin95d,Weinberg97,MSW,GV,GLV03}.
The growth of density fluctuations leading to galaxy formation
effectively stops when the vacuum energy dominates the Universe.  In
regions where $\rL$ is greater, it will dominate earlier, and thus
there will be fewer galaxies (and therefore fewer observers).  Suppose
${\tilde \rL}$ is the value for which the vacuum energy dominates at
about the epoch of galaxy formation, $z_\L\sim z_G$.  Then, values
$\rL\gg{\tilde\rL}$ will be rarely observed, because the density of
galaxies in the corresponding regions is very low.  On the other hand,
values $\rL\ll{\tilde\rL}$ are also rather unlikely, simply because this
range of values is very small. (This argument assumes that the
prior distribution for $\rL$ is not sharply peaked at $\rL=0$; see
below.)  Thus, a typical observer should expect to measure
$\rL\sim{\tilde\rL}$: the vacuum energy should dominate roughly at the
epoch of galaxy formation.

This anthropic prediction was subsequently confirmed by supernovae and
CMB observations.  The observed value \cite{supernovae,Spergel03} for
the vacuum energy density, in units of the critical density,
$\Omega_\L^{obs}\approx 0.7$, is in a qualitative agreement with the
anthropic probability distribution, which is peaked at
\cite{Vilenkin95a,Efstathiou95,MSW,GV} $\Omega_\L \sim 0.9$.  The most
recent analysis \cite{GLV03} finds that anthropic predictions are
in agreement with observations at a $2\sigma$ level. 

Here we show that the agreement can be further improved by allowing
for a nonzero neutrino mass. The value $\tilde \rho_\L$ at the peak of
the probability distribution for $\rho_\L$ depends on the magnitude of
density fluctuations on the galactic scale, $\sigma_G$. The
calculation in \cite{GLV03} was performed assuming massless neutrinos
and a scale-invariant spectrum of fluctuations, normalized using the
large-scale CMB measurements. If $\sigma_G$ is decreased, galaxies
form later, and $\tilde \rho_\L$ becomes smaller, thus improving the
agreement with $\rho_\L^{obs}$.  This is precisely the effect of
massive neutrinos, since they slow down the growth of density
fluctuations.  If the sum of the three neutrino masses is $m_\nu \sim
1-2$ eV, this suppression is sufficiently large to bring the anthropic
prediction for $\rho_\L$ into agreement with observations at the
$1\sigma$ level.

It is very intriguing that the value $m_\nu\sim 1$ eV for the sum of
the neutrino masses was itself recently predicted from anthropic
considerations.  It was suggested in \cite{TVP03} that the neutrino
masses might also be stochastic variables, changing from one part of
the Universe to another, and that their small values might be due to
anthropic selection.  In regions with a larger value of $m_\nu$, the
suppression of density fluctuations is stronger.  As a result, there
are fewer galaxies in such regions, and the corresponding values of
$m_\nu$ are less likely to be observed.  The probability distribution
for $m_\nu$ calculated in \cite{TVP03} is peaked at $m_\nu\approx 2$
eV, with a $1\sigma$ range between 0.5 eV and 3 eV. This 
corresponds to a mild but non-negligible suppression of
galaxy formation.\footnote{It is conceivable that the suppression of
galactic-scale density fluctuations relative to the largest observable
scales is not due to massive neutrinos, but reflects some feature in
the primordial fluctuation spectrum. 
Such suppressed small-scale fluctuations
have been suggested by certain
galaxy cluster observations
(assuming that the matter density parameter is $\Omega_M\sim 0.3$, see
\cite{Neta}),
and by the WMAP team \cite{Spergel03} who pointed
out that an initial spectrum with a decreasing running spectral index
is somewhat preferred by the data.
Whatever the origin of such a suppression, it
improves the agreement of the anthropic predictions for $\rL$ with
observations.}  
Indeed, recent
analysis of X-ray cluster data combined with CMB and 2dF
has suggested that $m_\nu=0.64^{+0.40}_{-0.28}$ eV \cite{Allen03},
although this remains controversial given  
cluster-related systematic uncertainties \cite{MaxSDSS03}. 
In this paper, we shall also consider models where both $m_\nu$ and $\rL$
are assumed to be stochastic variables.  

The paper is organized as follows. In Section \ref{sec:general} we
describe our general approach and justify our choice of the
theoretical prior distributions for $m_\nu$ and $\rho_\L$. In Section
\ref{sec:pmurho} we calculate the probability distribution for $\rL$
at nonzero values of $m_\nu$, as well as the joint probability
distribution for $m_\nu$ and $\rho_\L$ in models where both are
variable. We conclude with a discussion of our results in Section
\ref{sec:summary}. Some technical details are given in the Appendix.

\section{The method and the priors}
\label{sec:general}

The method used in this paper follows closely that in \cite{GV,TVP03}.
Consider a model in which $m_\nu$ and
$\rho_\L$ are allowed to vary from one part of the Universe to another. 
We define the probability
distribution ${\cal P}(m_\nu,\rho_\L) dm_\nu d\rho_\L$ as being proportional 
to the number of observers in the Universe who will measure $m_\nu$ in the
interval $dm_\nu$ and $\rho_\L$ in the interval $d\rho_\L$.  
This distribution can be represented as
\cite{Vilenkin95a,GV,TVP03} 
\ba
{\cal P}(m_\nu,\rho_\L) & \propto & {\cal
P}_{prior}(m_\nu,\rho_\L) \nonumber \\
&\times& \int dM \ n_G(m_\nu,\rho_\L,M) \ N_{obs}(m_\nu,\rho_\L,M).
\label{P}
\ea 
Here, ${\cal P}_{prior}(m_\nu,\rho_\L)dm_\nu d\rho_\L$ is the prior
distribution,  
which is proportional to the comoving volume of those parts of the Universe
where $m_\nu$ and $\rho_\L$ take values in intervals $dm_\nu$ and $d\rho_\L$, 
$n_G(m_\nu,\rho_\L,M)dM$ is the comoving 
number density of galaxies of mass in the 
interval $dM$ that will ever form in regions with given values of $m_\nu$ 
and $\rho_\L$, and $N_{obs}(m_\nu,\rho_\L,M)$ is the average 
number of observers per 
galaxy.  The distribution (\ref{P}) gives the probability that a
randomly selected 
observer is located in a region where the sum of the three neutrino
masses is in the interval $dm_\nu$ and the vacuum energy density is in the
interval $d\rho_\L$.

We shall assume, as it was done in \cite{MSW,GV,TVP03}, that the integral
in (\ref{P}) is dominated by 
large galaxies like the Milky Way, with
mass $M\gtrsim M_G\sim 10^{12}M_\odot$. We shall also assume, as a
rough approximation, that for
galaxies in this mass range $N_{obs}$ does not depend significantly
on $m_\nu$ or $\rho_L$ and is simply proportional to the number of stars, or to
the mass of the galaxy,
\be
N_{obs}(m_\nu,\rho_\L,M)\propto M \ .
\label{Nobs}
\ee 
(We shall comment on the validity of this assumption in Section \ref{sec:summary}.)
Then it follows from Eq.~(\ref{P}) that
\beq
{\cal P}(m_\nu,\rho_\L)\propto {\cal P}_{prior}(m_\nu,\rho_\L)F(M>M_G,m_\nu,\rho_\L),
\label{P'}
\eeq 
where $F(M>M_G,m_\nu,\rho_\L)$ is the fraction of matter that clusters into
objects of mass larger than $M_G$ in regions with these values of
$m_\nu$ and $\rho_\L$. 


The fraction of collapsed matter $F(M>M_G,m_\nu,\rho_\L)$ can be approximated
using the standard Press-Schechter formalism \cite{PressSchechter}.
We assume a Gaussian density fluctuation field $\delta({\bf x},t,m_\nu,\rho_\L)$
with a variance $\sigma(t,m_\nu,\rho_\L)$ on the galactic scale
$(M_G=10^{12}M_\odot)$,
\beq
P(\delta,t,m_\nu,\rho_\L)\propto \exp\left[-{\delta^2\over{2\sigma^2(t,m_\nu,\rho_\L)}}\right].  
\eeq 
At early times, when neutrinos are still relativistic, the variance $\sigma$
is assumed to be independent of $m_\nu$ and $\rho_\L$.
An overdense region collapses when the linearized density contrast
$\delta$ exceeds a critical value $\delta_c$, which in the spherical
collapse model with massless neutrinos is estimated to be
\cite{Weinberg87} $\delta_c\approx 1.63$. We shall assume that the
same value applies when neutrinos have a small mass.

In the Press-Schechter approximation, the asymptotic mass fraction in
galaxies of mass $M>M_G$ is
\ba
F(M>M_G,m_\nu,\rho_\L) &\propto& 
\erfc\left[{X\over\sqrt{2}}\right],
\label{nG}
\ea
where
\beq
X=\delta_c/{\sigma_\infty(m_\nu,\rho_\L)}
\label{X}
\eeq
and $\sigma_\infty (m_\nu,\rho_\L)\equiv \sigma(t\to\infty,m_\nu,\rho_\L)$.

\subsection{Prior distributions}
\label{subsec:priors}

The calculation of the prior distribution ${\cal
P}_{prior}(m_\nu,\rho_L)$ requires a particle physics model which
allows the neutrino masses and the dark energy density to vary and a
cosmological ``multiverse'' model that would generate an ensemble of
sub-universes with different values of $m_\nu$ and $\rho_\L$. 
Let us first focus on the neutrino masses. (The discussion in this
subsection follows \cite{DV,TVP03}.)

Dirac-type neutrino masses can be generated
if the Standard Model neutrinos $\nu^\alpha$ mix through the Higgs
doublet VEV $\Phi$ to some gauge-singlet fermions $\nu_c^\beta$,
\beq
g_{\alpha\beta}\Phi{\bar\nu}^\alpha\nu_c^\beta.
\label{Dirac}
\eeq
The couplings $g_{\alpha\beta}$ will generally be variable in string
theory inspired models involving antisymmetric form fields $F_a$
interacting with branes. (Here, the index $a$ labels different form
fields.) $F_a$ changes its value by $\Delta F_a=q_a$ across a brane,
where $q_a$ is the brane charge. In the low-energy effective theory,
the Yukawa couplings $g_{\alpha\beta}$ become functions of the form
fields,
\beq
g_{\alpha\beta}=\sum g_{\alpha\beta}^a
{{(F_a - F_a^{(0)})}\over{M_p^2}} + ... 
\label{g}
\eeq
Here, the summation is over all form fields, the coefficients
$g_{\alpha\beta}^a$ are assumed to be 
numbers $\sim 1$, and $M_p$ is the effective cutoff scale, which we 
assume to be the Planck mass. We have assumed that $g_{\alpha\beta}$
vanish at some point $F_a=F_a^{(0)}$ in the $F$-space; then
Eq.~(\ref{g}) is an expansion about that point.

In such models, closed brane bubbles nucleate and expand during
inflation \cite{Brown}, creating exponentially large regions with 
different values
of the neutrino masses. When $F_a$ changes in increments of $q_a$,
$m_\nu$ changes in increments of $\Delta m_\nu\sim \Phi q_a/M_p^2$. To
be able to account for neutrino masses $\lesssim 1$ eV, we have to
require that $\Delta m_\nu\lesssim 1$ eV, that is, 
\beq
q_a\lesssim 10^{-11} M_p^2, 
\label{smallq}
\eeq
for at least some of the brane charges. Such small values of the
charges can be achieved using the mechanisms discussed in
\cite{DV01,Banks,Feng}. 

The natural range of variation of $F_a$ in Eq.~(\ref{g}) is the Planck
scale, and the corresponding range of the neutrino masses is $0\leq
m_\nu^{(i)}\lesssim\Phi$. (Here, the index $i$ labels the three
neutrino mass matrix eigenvalues.) Only a small fraction of this range
corresponds to values of anthropic interest, $m_\nu\lesssim 10$~eV.
In this narrow anthropic range, we expect that the probability
distribution for $F_a$ after inflation is nearly flat \cite{GV01},
\beq
d{\cal P}_{prior} \propto dF_1 dF_2 ... ,
\eeq
and that the functions $g_{\alpha\beta}(F_a)$ are well
approximated by linear functions (\ref{g}). If all three neutrino masses vary
independently, this implies that
\beq
d{\cal P}_{prior} \propto
dm_\nu^{(e)}dm_\nu^{(\mu)}dm_\nu^{(\tau)}.
\eeq
The probability for the combined mass $m_\nu=\sum m_\nu^{(i)}$ 
to be between $m_\nu$ and $m_\nu+dm_\nu$ is then proportional to the 
volume of the triangular slab of thickness $\sim dm_\nu$ in the
3-dimensional mass space,
\beq
d{\cal P}_{prior}\propto m_\nu^2 dm_\nu.
\label{n=2}
\eeq
Alternatively, the neutrino masses can be related to one another, for
example, by a spontaneously broken family symmetry.  If all three
masses are proportional to a single variable mass parameter, then we
expect
\beq
d{\cal P}_{prior}\propto  dm_\nu.
\label{n=0}
\eeq

To reiterate, the prior distributions (\ref{n=2}),(\ref{n=0}) are
generic if the following two conditions are satisfied: (i) the full
range of variation of $m_\nu$ is much greater than the anthropic range
of interest and (ii) the point $m_\nu = 0$ is not a singular point of
the distribution. The values of $F_a$ which make the Yukawa couplings
(\ref{g}) vanish are not in any way special from the point of view of
the distribution for $F_a$, so there is no reason to expect a
singularity at that point, and thus the condition (ii) is likely to be
satisfied. 

As was shown in \cite{TVP03}, the
anthropic prediction for the neutrino masses with the prior given 
by Eq.~(\ref{n=2}) is in a strong disagreement with current
observational bounds. Allowing for a varying $\rho_\L$ is unlikely
to change this result.

There is also a possibility that the right-handed neutrinos
$\nu_c^\beta$ in (\ref{Dirac}) have a large Majorana mass $M_R\gg
\Phi$. In this case, small neutrino masses can be generated through
the see-saw mechanism, 
\beq
m_\nu\sim g^2\Phi^2/M_R. 
\eeq
If $M_R$ is variable, say, within a range $M_R\lesssim M_p$, then its
most probable values are likely to be $\sim M_p$, and the prior
distribution will favor $m_\nu\sim 10^{-6}$~eV. Such extremely small
values of $m_\nu$ are in conflict with the neutrino oscillation data.
In this paper we shall assume Dirac masses with the
prior (\ref{n=0}).

It should be noted that the Higgs potential and the Higgs expectation
value $\Phi$ in (\ref{Dirac}) will generally be functions of
$F_a$. Moreover, each field $F_a$ contributes a term $F_a^2/2$ to the
vacuum energy density $\rho_\L$, and regions with different values of
$F_a$ will generally have different values of $\rho_\L$. However, in
the presence of several form fields with sufficiently small charges,
variations of all these parameters are not necessarily correlated, and
here we shall assume that there is enough form fields to allow
independent variation of the relevant parameters. We can then consider
sub-ensembles of regions where some of the parameters are variable,
while the other are fixed.

In this paper we will be concerned with two models: (i) variable $\rL$
at a fixed value of $m_\nu$ and (ii) variable $m_\nu$ and $\rho_\L$,
all other parameters being fixed. A line of reasoning similar to that
above suggests that, in the anthropically interesting range, the prior
distribution for $\rho_\L$ should also be flat
\cite{Vilenkin95d,Weinberg97}
\beq
d{\cal P}_{prior}\propto d\rho_\L.
\label{priorL}
\eeq
Assuming independent variation of $m_\nu$ and $\rho_\L$, we then have
\beq
{\cal P}_{prior}(m_\nu,\rho_\L) = {\rm const}.
\label{priorn}
\eeq

We finally comment on the recent work \cite{Bousso,Susskind03,Douglas}
suggesting that string theory involves a large number of form fields
$F_a$ ($\sim 10^3$) and therefore admits an incredibly large number of
vacua ($\sim 10^{1000}$) characterized by different particle
physics parameters.\footnote{Apart from the form fields, the string
vacua are characterized by a number of other parameters specifying the
string compactifications.} The spectrum of $\rho_\L$ and $m_\nu$
could then be very dense even if the brane charges are not small,
$q_a\sim M_p^2$. However, in this ``discretuum'' of vacua, nearby
values of $m_\nu$ or $\rho_\L$ correspond to very different values of
the form fields, and we can no longer argue that the prior
distribution should be flat in the anthropic range. Calculation of the
prior in such a discretuum remains an important problem for
future research. In this paper we shall use the distribution
(\ref{priorn}).

\subsection{The full distribution}

The full probability distribution can now be obtained from Eqs.~(\ref{P'}),
(\ref{nG}), and (\ref{priorn}),
\be
{\cal P}(m_\nu,\rho_\L)\propto \erfc\left[{X\over\sqrt{2}}\right]\ ,
\label{P1}
\ee
where $X$ is given by Eq.~(\ref{X}),
\beq
X=\delta_c/{\sigma_\infty}.
\label{X1}
\eeq
Depending on the model, the asymptotic density contrast
$\sigma_\infty$ is a function of $m_\nu$ and $\rho_\L$ or of $m_\nu$
only.

\section{The probability distribution}
\label{sec:pmurho}

Let us consider models where both $\rho_\L$ and $m_\nu$ are allowed to vary
from one part of the Universe to another.
We shall calculate the probability distribution expressed in
terms of $\rho_\L$. We will also state some of our results in terms of
\be
\Omega_\L = (8\pi G/3)\rho_\L h^{-2} \equiv \omega_\L h^{-2} \ .
\ee 

To describe the growth of density fluctuations, it will be convenient
to use the ratio
\be
x\equiv {\rho_\L \over \rho_M(t)} = {\Omega_\L \over \Omega_M} (1+z)^{-3} \ 
\label{x}
\ee
as our time variable. The variance $\sigma_\infty(m_\nu,\rho_\L)$ 
in Eq.~(\ref{X1}) is proportional to the total linear growth factor of
the density perturbations, 
\be
D^{\infty}(f_\nu,\rho_\L) \equiv D(\fn,\rho_\L,x \rightarrow \infty) \ ,
\ee
where 
\be
f_\nu \equiv { \Omega_\nu \over \Omega_B+\Omega_{cdm}+\Omega_\nu}
\ee 
is the neutrino fraction of the total matter density.
Hence, the effect of neutrino masses on the fraction 
$F(M>M_G,m_\nu,\rho_\L)$ in (\ref{P1}) can be
taken into account by writing 
\be
\sigma_\infty(m_\nu,\rho_\L)
=\sigma_\infty(0,\rho_\L) {D^{\infty}(f_\nu,\rho_\L)\over D^{\infty}(0,\rho_\L)} \ ,
\label{sigma1a}
\ee
where 
$D^{\infty}(0,\rho_\L)$ is the total growth factor in regions with
$\fn=0$. 

The linear growth factor as a function of redshift can be found 
numerically, using standard codes such as CMBFAST \cite{cmbfast},
or analytically, using, e.~g. fitting formulae from \cite{EisensteinHu99}.
We have tested that the two methods give results that agree within a 
margin of error of a few percent. Technical details related to their 
evaluation can be 
found in the Appendix. Throughout this paper we choose to work with the 
semi-analytical fitting formulae given in Sec.~\ref{analgrowth}. 

We need to relate the asymptotic density contrast
$\sigma_\infty^{(0)}$ in Eq.~(\ref{sigma1a}) to the locally observable
quantities.  Current CMB measurements do not extend down to the
galactic scale, and to sidestep issues related to light-to-mass bias,
the galactic density fluctuations are usually
inferred by extrapolating from larger scales (on which the effect of
neutrino free streaming is unimportant) using a nearly scale-invariant
fluctuation spectrum and assuming that the neutrino masses are
negligible. The growth factor evaluated in this way overlooks the
non-zero massive neutrino fraction $\fn^*$ in our part of the
Universe, which is currently unknown but potentially significant.
This growth factor, ${\hat D}(\fn^*,\rho^*_\L,x)$, is evaluated under the
assumption that neutrinos are massless, but with $\Omega_\nu^*$ added
to $\Omega_{cdm}$.  Here and below, hats indicate the quantities
evaluated for our local region assuming massless neutrinos, and
superscript ``*'' indicates the actual local values of these
quantities. For example, $x^*\approx 2.3$ 
is the present local value of
$x$ and ${\hat\sigma}$ is the asymptotic density contrast in our
region calculated assuming massless neutrinos.  Hence, we can write
\be
\sigma_\infty(0)
=\hat \sigma_\infty {D^{\infty}(0,\rho_\L)\over 
{\hat D}^{\infty}(\fn^*,\rho_\L^*)} \ ,
\label{sigma2a}
\ee
where ${\hat D}^{\infty}(\fn^*,\rho^*_\L) 
\equiv {\hat D}(\fn^*,\rho_\L^*,x\rightarrow \infty)$.
 
To evaluate ${\hat\sigma}_\infty$, we again note that in the absence of massive
neutrinos, the growing mode of density fluctuations is proportional to 
$D(0,\rho_\L,x)$. Therefore, the asymptotic density contrast $\hat \sigma_\infty$ is 
related to the present value ${\hat\sigma}(M)$ via
\ba
\hat \sigma_\infty &=& \hat \sigma(M) {{\hat D}^{\infty}(\fn^*,\rho^*_\L)
\over {\hat D}(f^*_\nu,\rho^*_\L,x^*)} \ .
\label{sigma3a}
\ea

Finally, using Eqs. (\ref{sigma1a}), (\ref{sigma2a}) and (\ref{sigma3a}), 
we can rewrite Eq.~(\ref{X1}) as
\ba
X 
={\delta_c \over \hat \sigma(M)} 
{ {\hat D}(\fn^*,\rho^*_\L,x^*) \over D^{\infty}(f_\nu,\rho_\L)} \ .
\label{X2}
\ea
Here, ${\hat\sigma}(M)$ is the density contrast on the
galactic scale $M=10^{12}M_{\odot}$ inferred from the large-scale CMB
data assuming massless neutrinos.  We emphasize that the {\it actual}
density contrast should be smaller if indeed the suppression of the
galactic density fluctuations due to neutrinos is
non-negligible.\footnote{When better data for the actual linearized
galactic-scale density contrast is available, it can also be used for
the evaluation of $F(M>M_G,m_\nu,\rho_\L)$. The only modifications needed in
Eq.~(\ref{X2}) are that ${\hat\sigma}(M)$ has to be replaced by
$\sigma^*(M)$ and the factor ${\hat D}(\fn^*,\rho_\L^*,x^*)$ should be replaced 
by $D(\fn^*,\rho_\L^*,x^*)$.}. To estimate
${\hat\sigma}(M)$ we can use the linear power spectrum inferred from
measurements of CMB anisotropies. First we need to find the length scale
$R_G$ corresponding to the mass scale $M=10^{12}M_{\odot}$. It can be
found from
\ba
R(M) &=& \left({3M \over 4\pi \rho_0}\right)^{1/3} \nonumber \\
&\approx& 0.951 \ \Omega_M^{-1.3} h^{-2/3}
\left( {M \over 10^{12}M_{\odot}} \right)^{1/3} {\rm Mpc} \ ,
\ea
where $\rho_0 \approx 1.88 \cdot 10^{-26} \ {\Omega_M h^2} \ {\rm kg/m}^3$ is the mean
cosmic density. For $h=0.72$ and $\Omega_M =0.27$ this gives
\be
R_G \approx 1.32 \ {\rm h^{-1} Mpc} \ .
\ee
The corresponding
linearized density contrast ${\hat \sigma}_{R_G}$ found using WMAP's best fit power 
law model \cite{Spergel03} is
\be
{\hat \sigma}_{R_G} \approx {\hat \sigma}(10^{12} M_{\odot}) 
\approx 2.41 \pm 0.26 \ .
\label{sigmarg}
\ee

Substituting analytical expressions for ${\hat D}(\fn^*,\rho_\L^*,x^*)$ 
and $D^{\infty}(f_\nu,\rho_\L)$, derived in the Appendix (Sec.~\ref{analgrowth}),
into Eq.~(\ref{X2}) and using  the fact that $x^* \gg x_{eq}$ gives
\ba
X &=& {\delta_c \over \hat \sigma(M)} \left({3\over 2} 
(x_{eq}^*)^{-1/3} \right)^{1-p} \nonumber \\ &\times&
\left( {1-\fn \over 1-\fn^*} \right)^{4p/3} 
\left({\omega_\Lambda \over \omega_\Lambda^*}\right)^{p/3}
{ G(x^*) \over [G(\infty)]^p} \ ,
\label{X2b}
\ea
where 
\beq
p={1\over{4}}\left(\sqrt{25-24f_\nu}-1\right)
\eeq
and $G(x)$ is the growth factor in a flat universe filled with 
pressureless matter and vacuum energy, given by \cite{heath77,MSW}:
\be
G(x)={5\over 6}\left[ {1+x \over x}\right]^{1/2} 
\int_0^x{dw \over w^{1/6} (1+w)^{3/2}} \ ,
\label{fx}
\ee
with the asymptotic value $G(\infty)\approx 1.44$.

The probability distribution ${\cal P}(m_\nu,\rho_\L)$ is obtained 
from Eqs.~(\ref{P1}) and (\ref{X2}) and depends on parameters
$\rho^*$, ${\hat \sigma}_{R_G}$ and $f_\nu^*$. The last parameter,
the value of $f_\nu$ in our part of the Universe, is one of the
quantities we are trying to predict. On the other hand, we found that 
the distribution is not very sensitive to the assumed value 
$m_\nu^*$: varying $m_\nu^*$ between 0.1 and 3 eV changes the location of
the peak of distribution functions by no more than $10\%$ percent.  
This uncertainty can be accounted for by
multiplying ${\cal P}(m_\nu,\rho_\L)$ by an appropriately chosen prior
probability ${\tilde p}(m^*_\nu)$, reflecting the current state of our
knowledge of the value of $m_\nu^*$, and integrating over $m^*_\nu$.
Current $95$\% upper bound on the sum of neutrino masses, obtained
using the WMAP and SDSS data \cite{MaxSDSS03}, is $m^*_\nu \lesssim
1.7$ eV.  Hence, we choose ${\tilde p}(m^*_\nu)$ to be a Gaussian
centered at $m^*_\nu=0$ with a standard deviation of $\sigma_{m^*_\nu}
= 0.85 eV$. 
A similar procedure is also needed to account for the uncertainty in
our knowledge of $\rho_\L^*$
and ${\hat \sigma}_{R_G}$.
We compute $\omega_\L^*=\Omega_\Lambda^* {h^*}^2$ using
$h^*=\bar h \pm \sigma_h=0.72\pm 0.08$ measured by the HST project
\cite{hst}
and  $\Omega_\Lambda^*=\bar \Omega_\L \pm \sigma_{\Omega_\Lambda} =
0.7\pm 0.08$,
a conservative estimate from CMB and large-scale structure observations
\cite{Spergel03,MaxSDSS03}.
We take ${\tilde p}(\omega_\Lambda^*)$ to be a Gaussian centered at 
$\bar \omega_\L^*= {\bar \Omega_\L} {\bar h}^2$, 
with a standard deviation
$\sigma_{\omega_\Lambda} = \bar \Omega_\Lambda 2 \bar h \sigma_h +
\bar h  
\sigma_{\Omega_\Lambda}$. We account for the uncertainty in
${\hat \sigma}_{R_G}$ taking ${\tilde p}({\hat \sigma}_{R_G})$ to be
a Gaussian with the mean and the standard deviation given by Eq.~(\ref{sigmarg}).
Hence, we have
\ba
{\cal P}(m_\nu,\rho_\Lambda) &\rightarrow&
\int d m_\nu^* {\tilde p}(m_\nu^*) 
\int d \omega_\Lambda^* {\tilde p}(\omega_\Lambda^*) \nonumber \\ &\times&
\int d {\hat \sigma}_{R_G} {\tilde p}({\hat \sigma}_{R_G}) \ 
{\cal P}(m_\nu,\rho_\Lambda) \ .
\label{marginalize}
\ea

\begin{figure}[tb]
\centerline{\epsfxsize=8.0cm\epsffile{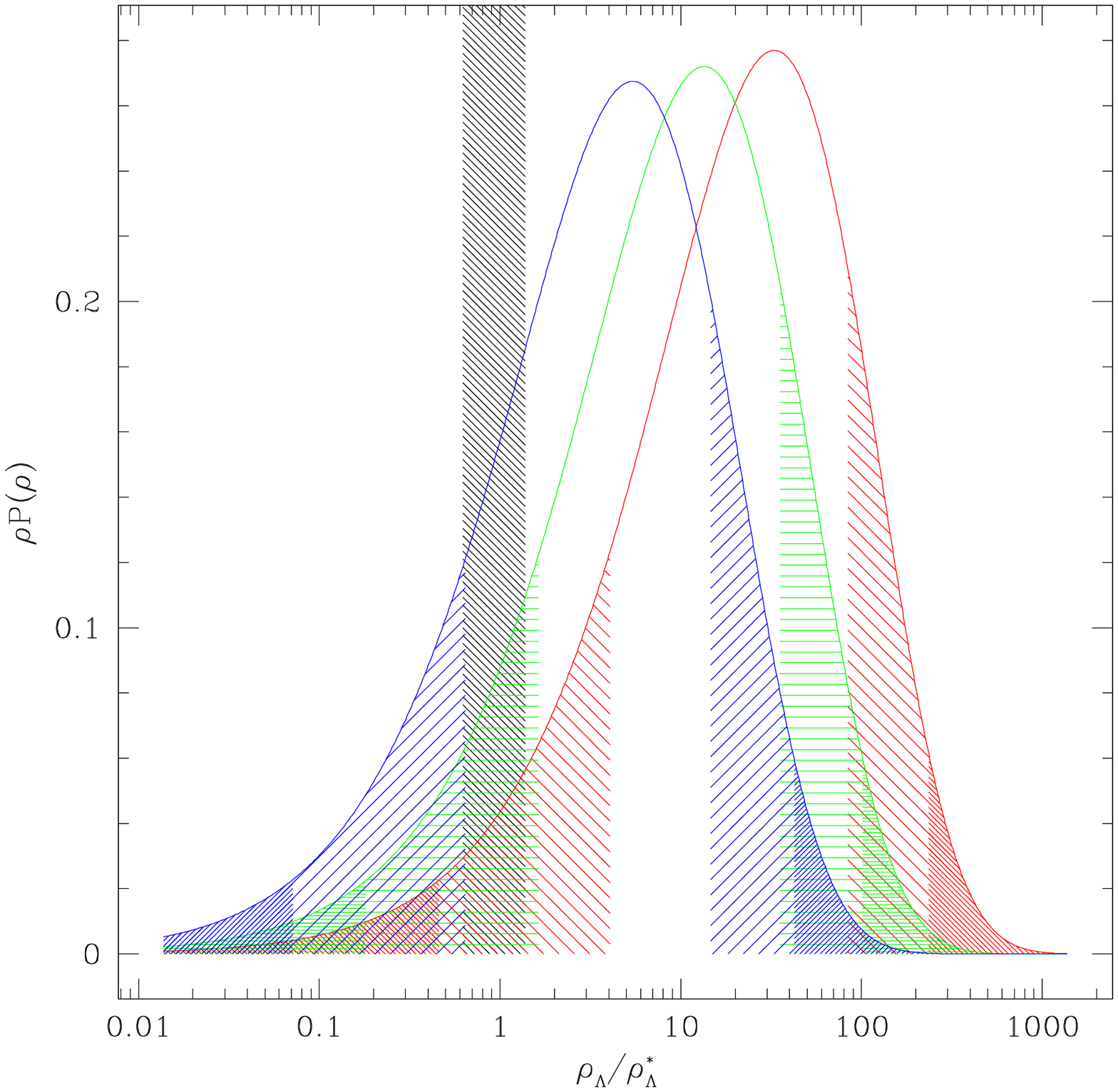}} 
\caption{ \label{prho3mu}\footnotesize 
The logarithmic probability distributions $\rho_\L{\cal P}(m_\nu,\rho_\L)$ vs
$\rho_\L/\rho_\L^*$
for three fixed values of $m_\nu$: $0$, $1$ and $2$ eV, peaked from right to left.
In each of the three cases, the lightly and densely shaded areas 
are the regions excluded at $68$\% and $95$\% level respectively.
Also shown (the dark vertical strip), is the $1\sigma$ uncertainty in 
$\rho_\L^*$ derived from observations.}
\end{figure}

In the following two subsections we consider two cases. In the first,
we calculate the probability density for $\rho_\L$, while using
several fixed values of $m_\nu$. In the second, we let both
$\rho_\L$ and $m_\nu$ vary and calculate their joint probability
as well as the effective probability for each of the two parameters.

\subsection{$P(\rho_\L)$ for fixed values of $m_\nu$}
In this subsection we consider a few 
fixed values of $m_\nu$ and evaluate corresponding probabilities
for $\rho_\L$.

The probability ${\cal P}(\rho_\L)$ per log~$\rho_\L$ for
$m_\nu=0$, $1$ and $2$ eV is shown in Fig.~\ref{prho3mu} along with
the $1$ and $2\sigma$ bounds in the three cases. These bounds are
found by cutting off equal measures of probability on both ends
($16$\% and $2.5$\% for $1$ and $2\sigma$, respectively).  Also shown
in the figure is the observationally favored range of $\rho_\L$. As
one can see from the figure, the observed value is within the
$2\sigma$ range of the anthropic prediction in all three cases and
within the $1\sigma$ bound in the $m_\nu=2$eV case.

\subsection{Joint probability $P(m_\nu,\rho_\L)$}

Let us now calculate the probability distribution ${\cal
P}(m_\nu,\rho_\L)$ with both $m_\nu$ and $\rho_\L$ variable.

\begin{figure}[t]
\centerline{\epsfxsize=8.0cm\epsffile{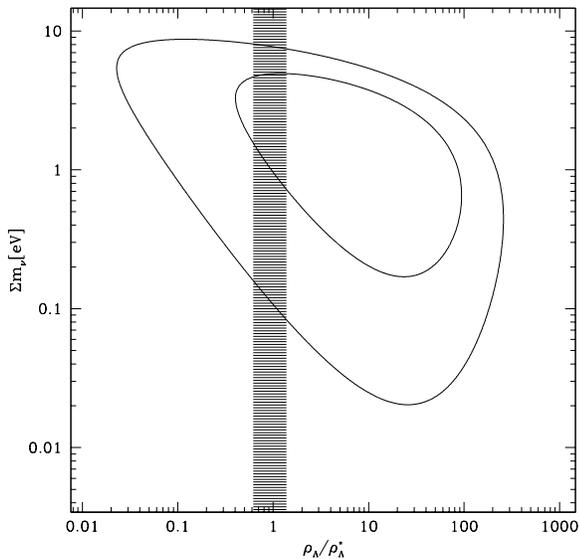}}
\caption{ \label{cont2}\footnotesize 
The $68\%$ and $95\%$ probability contours in the
$m_\nu$-$\rho_\Lambda$ parameter plane
for flat priors. The shaded vertical strip shows the $1\sigma$
uncertainty in the observed value of $\rho_\Lambda$.}
\end{figure}

We are interested in finding the 
$68\%$ and $95\%$ probability contours in the $m_\nu$-$\rho_\Lambda$
parameter plane. The simple procedure of cutting off equal measures of 
probability on both ends used in the previous subsection does not
readily generalize to the case of more than one variable. 
Instead, we find contours of constant 
logarithmic probability density
$m_\nu \rho_\L {\cal P}(m_\nu,\rho_\L)$
that bound $68\%$ and $95\%$ of the volume under the surface defined by the
probability distribution ${\cal P}(m_\nu,\rho_\L)$.
These contours are shown in Fig.~\ref{cont2}.
\begin{figure}[tb]
\centerline{\epsfxsize=8.0cm\epsffile{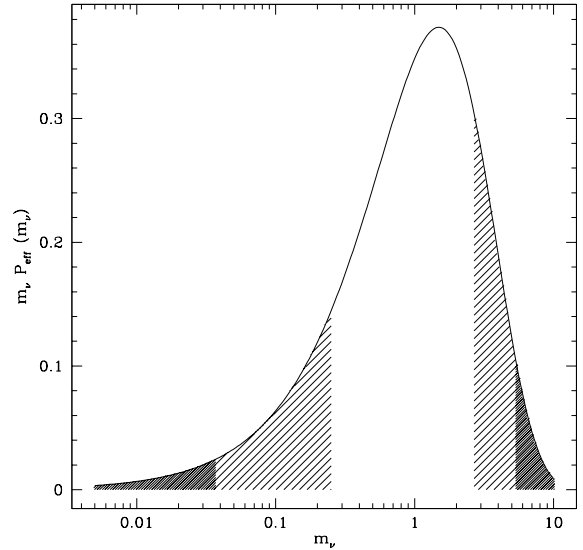}}
\caption{\label{mnub} 
Effective logarithmic probability distribution for $m_\nu$ 
and the boundaries of the $68\%$ and $95\%$ probability regions.}
\end{figure}
The ensemble averaged values of the dark energy density and the neutrino masses found
using the joint probability distribution ${\cal P}(m_\nu,\rho_\L)$ are
\be
\langle \Omega_\Lambda \rangle =0.85 \ , \ 
\langle \rho_\Lambda/\rho_\Lambda^* \rangle = 21 \ , \
\langle  m_\nu \rangle = 1.5 {\rm eV} \ .
\ee
\begin{figure}[tb]
\centerline{\epsfxsize=8.0cm\epsffile{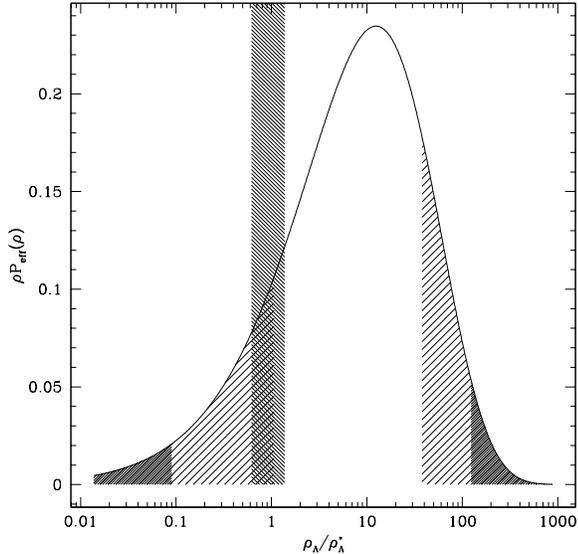}}
\caption{\label{rhob} 
The effective probability from Eq.~(\ref{peffL}) plotted per log $\rho_\L/\rho_\L^*$.
The lightly and densly shaded regions are excluded at $1$ and $2\sigma$ level, respectively.
Expressed in terms of $\Omega_\L$, the $1\sigma$ region is given by $0.695 < \Omega_\L < 0.988$.}
\end{figure}

Next we find the effective one dimensional probability distributions
for $m_\nu$ and $\rho_\L$. The probability
distribution for $m_\nu$ is found by integrating ${\cal
P}(\rho_\L,m_\nu)$ over $\rho_\L$:
\be 
{\cal P}_{\rm eff}(m_\nu)=\int d\rho_\Lambda {\cal P}(\rho_\Lambda,m_\nu) \ .
\label{peffnu}
\ee
In Fig.~\ref{mnub} we show the plot of ${\cal P}_{\rm eff}(m_\nu)$ per log~$m_\nu$,
along with the boundaries of the $68\%$ and $95\%$ probability regions.

In Fig.~\ref{rhob} we show
\be 
{\cal P}_{\rm eff}(\rL)=\int dm_\nu {\cal
P}(\rL,m_\nu) \ ,
\label{peffL}
\ee
plotted per log~$\rho_\L$, with the $68\%$ and $95\%$ probability
regions. We find, as might be expected, that
allowing for variable neutrino masses has a significant effect on the
anthropic prediction for $\rL$. In particular, it results
in the currently observed value, $\Omega_\Lambda=0.7 \pm 0.08$, now being
within the $1\sigma$ allowed region.

We note that the integrated distributions
(\ref{peffnu}),(\ref{peffL}) should be interpreted with care. In
Eq.~(\ref{peffnu}), for example, the
integration is performed over the full range of $\rL$, which includes a
large range of values which are already observationally
excluded. Hence, this distribution cannot be used directly to make
observational predictions for $m_\nu$ in our region.

\section{Discussion}
\label{sec:summary}

In conclusion, we have argued that both the small vacuum energy density and
small values of the neutrino masses may be due to
anthropic selection. We have considered a model where $m_\nu$ and $\rho_\L$ 
vary independently, with flat priors for both variables. The predictions of
this model are qualitatively similar to those derived by considering
sub-ensembles with only one of these parameters variable and the other
fixed. But there is an interesting quantitative difference. The
probability distribution for the joint ensemble is shifted towards
smaller values of both $m_\nu$ and $\rho_\L$. As a consequence, the
agreement between the predicted and observed values of $\rho_\L$ is
improved.

In this work we have only considered the case of positive $\rho_\L$.
Negative values of $\rho_\L$ were discussed in \cite{GV} and in \cite{KL03}. There, it
was found that the probability of $\rho_\L < 0$ is comparable to that of 
$\rho_\L > 0$.

In our analysis we have assumed that the number of observers does 
not sensitively depend of $m_\nu$ or $\rho_\L$ and is simply proportional to the mass
of the galaxy, Eq.~(\ref{Nobs}). This appears to be a reasonable
assumption for small values of $m_\nu$ and $\rho_\L$, such that they
have little effect on structure formation. For larger values, the
properties of the galaxies, including the number of habitable stars
per unit mass, may be affected.  For example, if $\rho_\L$ gets large,
structure formation ends early and galaxies have a higher density of
matter. This may increase the danger of nearby supernova explosions
and the rate of near encounters with stars, large molecular clouds, or
dark matter clumps. Gravitational perturbations of planetary systems
in such encounters could send a rain of comets from the Oort-type
cloud towards the inner planets, causing mass extinctions. Further
increase of density may lead to frequent disruptions of planetary
orbits \cite{Q}. The effect of all this is to suppress the
probability density at large values of $\rho_\L$ and thus to shift the
distribution towards smaller values. This may further improve the
agreement of the anthropic predictions with the data. With better
understanding of galaxy and star formation, we may be able to obtain a
quantitative measure of these effects in not so distant future.

\bigskip

We thank Neta Bahcall, Gia Dvali and Jaume Garriga for
useful comments and discussions. This work was supported in part by 
the National Science Foundation.
AV was supported in part by the National Science
Foundation and the John Templeton Foundation.
MT was supported by NSF grant AST-0134999, 
NASA grant NAG5-11099, Research Corporation and the 
David and Lucile Packard Foundation. 

\appendix
\section{}
\label{appendixA}

To compare the growth of linear density fluctuations in models
with different values of $m_\nu$ and $\rho_\L$, we choose the
parameters so that all models have the same density contrast evolution 
while neutrinos are still relativistic and dark energy is negligible.
This is achieved by fixing the values of $\omega_c \equiv \Omega_{cdm} h^2$,
$\omega_b \equiv \Omega_b h^2$ and $T_{CMB}$ today. 
Then, for
given values of $\fn$ and $\rho_\Lambda=3\omega_\L/(8\pi G)$, we find
\be
h=\sqrt{{\omega_c+\omega_b \over 1-\fn} + \omega_\L}  \ ,
\ee
and obtain 
$\Omega_b$, $\Omega_{cdm}$, $\Omega_\Lambda$ and $\Omega_\nu$
using
\ba
\Omega_b = {\omega_b \over h^2} , \
\Omega_{cdm} = {\omega_c \over h^2} , \
\Omega_\Lambda = {\omega_\Lambda \over h^2} \ , \nonumber \\
\Omega_\nu = { f_\nu(\Omega_{cdm}+\Omega_b) \over 1 - f_\nu} \ .
\label{parameters}
\ea
We use the values $\omega_b=0.024$ and $\omega_c=0.132$, 
estimated from CMB and large-scale structure observations
\cite{Spergel03,MaxSDSS03}.

\subsection{Evaluating growth functions using CMBFAST}

The matter growth functions can be evaluated numerically, using
CMBFAST \cite{cmbfast}. Given the values of cosmological parameters,
including $\fn$ and $\rho_\L$, CMBFAST can compute Fourier components of the matter
density contrast, $\delta(f_\nu,\rho_\L,k,x)$, where $x$ is defined in
Eq.~(\ref{x}).
On scales much smaller than the neutrino free-streaming scale, $k\gg
k_{fs}$, neutrinos do not participate in gravitational clustering, so
one expects ratios such as $\delta(f_\nu',\rho_\L',k,x')/\delta(f_\nu,\rho_\L,k,x)$ to
be independent of $k$.  Galactic scales, corresponding to $k_g \sim
1{\rm Mpc}^{-1}$, satisfy this requirement for $m_\nu \ll 30$ eV, which
includes the range of interest to us
\footnote{In addition to this requirement, in order for this ratio of $\delta$s to
be independent of $k$, the scale under consideration had to 
enter the horizon sufficiently long before $z_{eq}$. Galactic scales, 
corresponding to $k_g \sim 1{\rm Mpc}^{-1}$ satisfy this constraint as well.}.
Therefore, since Eq.~(\ref{X2}) contains ratios of growth factors, we
can simply replace ${\hat D}(\fn^*,\rho_\L*,x^*)$ and $D^{\infty}(\fn,\rho_\L)$ by
${\hat \delta}(f^*_\nu,\rho_\L*,k_g,x^*)$ and $\delta(f_\nu,\rho_\L,k_g,x \rightarrow
\infty)$.  Namely, we write
\be
X={\delta_c \over \hat \sigma(M)} 
{{\hat \delta}(f^*_\nu,\rho_\L*,k_g,x^*) \over 
\delta(f_\nu,\rho_\L,k_g,x \rightarrow \infty)} \ ,
\label{X3}
\ee
where ${\hat \delta}(f^*_\nu,\rho_\L*,k_g,x^*)$ is the density contrast evaluated
assuming neutrinos are massless, but with $\Omega_\nu^*$ added to 
$\Omega_{cdm}$. We have checked that, as expected, the results are not 
sensitive to a particular choice of the value of $k$, provided that $k$ is
sufficiently large.

\subsection{Evaluating growth functions analytically}
\label{analgrowth}

In addition to numerical methods, there are analytical fitting
formulae describing the evolution of density fluctuations in the
presence of massive neutrinos. The growth of linear density
fluctuations on scales $k\gg k_{fs}$ in a Universe dominated by
nonrelativistic matter is given by \cite{Bond80}
\beq
\sigma(z)\propto (1+z)^{-p},
\eeq
where
\beq
p={1\over{4}}\left(\sqrt{25-24f_\nu}-1\right)\approx 1-{3\over{5}}f_\nu,
\label{p}
\eeq
and the last step in (\ref{p}) assumes
that $f_\nu\ll 1$.  The growth of fluctuations effectively begins at
the time of matter domination and terminates at vacuum
domination.

The effect of massive neutrinos on the growth factor can 
be calculated using \cite{EisensteinHu99}
\beq
D(\fn,\rho_\L,x)=[{\hat D}(\fn,\rho_\L,x)]^p.
\label{EH}
\eeq
The functional dependence on $\fn$ in Eq.~(\ref{EH})
is not accurate for very small values of $m_\nu$, when neutrinos become
nonrelativistic well after matter-radiation equality. But in this case
the $\fn$-dependence is very weak anyway. It has been verified 
in \cite{EisensteinHu99} that Eq.(\ref{EH}) is accurate to within $3\%$ 
in the whole relevant range of
neutrino masses. 

In a flat universe filled with pressureless matter and 
vacuum energy the growth factor, $G(x)$, is given by Eq.~(\ref{fx}).
For $x\ll 1$, it reduces to the
familiar growth function in a matter-dominated universe,
$G(x)\approx x^{1/3} \propto (1 + z)^{-1}$.

The effect of radiation can be included using the exact formula for a
matter plus radiation universe (in the absence of dark energy)
\cite{Peebles}, 
\beq
{\tilde G}(x) = 1+{3\over{2}}(x/x_{eq})^{1/3}.
\eeq
The density of radiation is negligible when the dark energy becomes
important, and vice versa. Hence, we can write, to a good accuracy,
\be
{\hat D}(\fn,x) = 1 + {3\over{2}}x_{eq}^{-1/3}(\fn) G(x) \ ,
\label{dx}
\ee
where
\be
x_{eq}^{-1/3}(\fn)= (x^*_{eq})^{-1/3} 
\left({1-\fn^* \over 1-\fn}\right)^{4/3} 
\left({\omega_\Lambda^* \over \omega_\Lambda}\right)^{1/3} 
\ .
\label{xeqfnu}
\ee
Note that the growth factor is normalized so that ${\hat D}(\fn,x\to 0)=1$.
Equation (\ref{xeqfnu}) follows from
\ba
x_{eq}^{-1/3} &=& (1 + z_{eq})/(1 + z_\L) 
\propto  \Omega_M^{4/3} h^2 \Omega_\L^{-1/3} \nonumber \\
&\propto& (1 - f_\nu)^{-4/3} (\omega_b + \omega_c)^{4/3} \omega_\L^{-1/3} \ ,
\ea
and, since $\omega_b$ and $\omega_c$ are fixed, we have
\be
x_{eq}^{-1/3}  \propto  (1 - f_\nu)^{-4/3} \omega_\Lambda^{-1/3} \ .
\ee
Eq.~(\ref{xeqfnu}) also depends on $x^*_{eq}$, currently estimated to
be 
\be
(x_{eq}^*)^{-1/3} \approx 2820 \ ,
\ee
and $\fn^*$, the value of which is currently unknown. As discussed in
Sec.~\ref{sec:pmurho}, the dependence of our results on $\fn^*$ is rather weak
and we opt to marginalize over it.

An approximation for $G(x)$ accurate for all values of $(x)$ 
was derived in \cite{TVP03}:
\be
G(x) \approx 
x^{1/3}\left[1+\left({x\over G(\infty)^3}\right)^\alpha\right]^{-1/3\alpha} \ ,
\ee
where $\alpha=0.795$.
At large $x$, the growth of density fluctuations is stalled, and 
$G(x)$ approaches the asymptotic value $G(\infty)\approx 1.44$.




\begin{thebibliography}{99}


\bibitem{Weinberg87}
\rf\nn Weinberg S;1987;Phys. Rev. Lett.;59;2607

\bibitem{Vilenkin95a}
\rf\nn Vilenkin A;1995;Phys. Rev. Lett.;74;846

\bibitem{Efstathiou95}
\rf\nn Efstathiou G;1995;MNRAS;274;L73

\bibitem{Vilenkin95d}
A. Vilenkin, in {Cosmological constant and the evolution of the
universe}, ed by K. Sato, T. Suginohara and N. Sugiyama (Universal
Academy Press, Tokyo, 1996); gr-qc/9512031.

\bibitem{Weinberg97}
S.~Weinberg, in ``Critical Dialogues in Cosmology'', 
proceedings of a Conference held at Princeton, New Jersey, 24-27 June 1996, 
Singapore: World Scientific, edited by Neil Turok, 1997., p.195.

\bibitem{MSW}
H. Martel, P.R. Shapiro and S. Weinberg, Ap. J. {\bf 492}, 29 (1998).

\bibitem{GV}
J. Garriga and A. Vilenkin, Phys. Rev. {\bf D67}, 043503 (2003).

\bibitem{GLV03}
J. Garriga, A.D. Linde and A. Vilenkin, hep-ph/0310034.

\bibitem{Davies}
P.C.W. Davies and S. Unwin, Proc. Roy. Soc. {\bf 377}, 147 (1981).

\bibitem{BT}
\rfbook\nnn Barrow J D\dualand \nnn Tipler F J;1986;The 
Anthropic Cosmological Principle;Clarendon Press;Oxford

\bibitem{LindeLambda}
A.D. Linde, in {\it 300 Years of Gravitation}, ed. by S.W. Hawking and
W. Israel, Cambridge University Press, Cambridge (1987).


\bibitem{supernovae} A.~Riess {\it et. al.}, Astron.~J.~{\bf 116}, 1009 (1998);
S.~Perlmutter {\it et. al.}, Astrophys.~J.~{\bf 517}, 565 (1999).

\bibitem{Spergel03}
\rfprep\nnn Spergel D N {\etal};2003;astro-ph/0302209 

\bibitem{TVP03} M.~Tegmark, A.~Vilenkin and L.~Pogosian, astro-ph/0304536.

\bibitem{Neta}
N. Bahcall {\it et. al.}, ApJ {\bf 585}, 182 (2003).

\bibitem{Allen03}
S.W. Allen, R.W. Schmidt and S.L. Bridle, MNRAS {\bf 346}, 593 (2003).

\bibitem{PressSchechter}
\rf\nnn Press W H\dualand\nn Schechter P;1974;ApJ;187;425

\bibitem{DV}
G. Dvali and A. Vilenkin, unpublished.

\bibitem{Brown}
J.D. Brown and C. Teitelboim, Nucl. Phys. {\bf 279}, 787 (1988).







\bibitem{DV01}
G. Dvali and A. Vilenkin, Phys. Rev. {\bf D64}, 063509 (2001);
hep-th/0304043.  

\bibitem{Banks}
T. Banks, M. Dine and L. Motl, JHEP 0101:031 (2001).


\bibitem{Feng}
J.L. Feng, J. March-Russell, S. Sethi and F. Wilczek, Nucl. Phys. {\bf
B602}, 307 (2001).

\bibitem{GV01}
J. Garriga and A. Vilenkin, Phys. Rev. {\bf D64}, 023517 (2001).

\bibitem{Bousso}
R. Bousso and J. Polchinski, JHEP 0006:006 (2000).

\bibitem{Susskind03}
\rfprep\nn Susskind L;2003;hep-th/0302219

\bibitem{Douglas}
S.~Ashok and M.~R.~Douglas, hep-th/0307049 .



















 



\bibitem{cmbfast} M.~Zaldariaga and U.~Seljak,
Astrophys.~J.~{\bf 469}, 437 (1996); http://www.cmbfast.org

\bibitem{EisensteinHu99}
\rf\nnn Eisenstein D J\dualand\nn Hu W;1999;ApJ;511;5

\bibitem{heath77}
\rf\nnn Heath D J;1977;MNRAS;179;351

\bibitem{hst} W.~L.~Freedman {\it et. al.} \ (HST Key Project),
Astrophys.~J.~{\bf 553}, 47 (2001)

\bibitem{MaxSDSS03} 
M.~Tegmark {\it et. al.} (SDSS Collaboration), astro-ph/0310723 .

\bibitem{KL03} R.~Kallosh and A.~Linde, Phys.~Rev.~{\bf D67}, 023510 (2003).

\bibitem{Q}
M. Tegmark and M.J. Rees, Ap. J. {\bf 499}, 526 (1998).

\bibitem{Bond80}
\rf\nnn Bond J R, \nn Efstathiou G\multiand\nn Silk, J;1980;PRL;45;1980

\bibitem{Peebles}
P.~J.~E.~Peebles,{\it Principles of Physical Cosmology}, Princeton University Press, 
Princeton, New Jersey (1993).


















\end{thebibliography}
\end{document}